\documentstyle[12pt]{article}
\newcommand\be{\begin{equation}}
\newcommand\ee{\end{equation}}
\newcommand\ba{\begin{eqnarray}}
\newcommand\ea{\end{eqnarray}}

\newcommand{\cl}{\centerline}
\newcommand\bear{\begin{eqnarray*}}
\newcommand\eear{\end{eqnarray*}}

\begin{document}
\setlength{\textwidth}{5.0in}
\setlength{\textheight}{7.5in}
\setlength{\parskip}{0.0in}
\setlength{\baselineskip}{18.2pt}
\hfill
{\tt HD-THEP-03-28}
\begin{center}
{\large{\bf Dirac's Constrained Hamiltonian Dynamics from an 
Unconstrained Dynamics}}
\vskip 0.3cm
\end{center}
\begin{center}
{Heinz J. Rothe}\par
\vskip 0.3cm
{Institut f\"ur Theoretische Physik}\par
{Universit\"at Heidelberg, Philosophenweg 16, D-69120 Heidelberg, Germany}
\footnote{email: h.rothe@thphys.uni-heidelberg.de}
\cl{\today}
\end{center}
\centerline{\bf Abstract}
We derive the Hamilton equations of motion for a constrained system in the 
form given by Dirac, by a limiting procedure, starting from the Lagrangean  for an unconstrained system. We thereby ellucidate the role played by the 
primary constraints and their persistance in time. 
\vskip1cm
The Hamiltonian formulation for systems whose dynamics is described by a Lagrangean with singular Hessian has been given a long time ago by Dirac 
\cite{Dirac}, and has been elaborated ever since in numerous papers \footnote{It is impossible to quote here the very large number or papers. Comprehensive 
discussions can be found e.g. in \cite{Teitelboim}}. As is well known all gauge theories fall 
into the class of singular systems. The usual starting point for deriving the Hamilton equations of motion is the singular Lagrangean which leads, in the language of Dirac, to primary constraints. While the primary constraints have no analog on the Lagrangean level, requiring their persistance in time 
leads to equations relating the coordinates and velocities. These are part of the Euler-Lagrange equations of motion. This 
persistance requirement may lead to secondary constraints and is the first step in the Dirac algorithm generating the constraints.
 The purpose of this paper is to derive the Hamilton equations of motion for systems with a singular Hessian, by starting from an unconstrained system and taking an appropriate limit. The formulation of the equations of motion within an extended phase space is quite natural, and the role played by the primary constraints, and the demand for 
their persistance in time will be illuminated thereby.

Consider the following Lagrangean quadratic in the velocities
\be\label{Lagrangean}
L = \frac{1}{2}\sum_{ij}W_{ij}(q;\alpha)\dot q_i\dot q_j - 
\sum_i\eta_i(q)\dot q_i - V(q) \ .
\ee
where $q$ stands for the set of coordinates $\{q_i\}$ and $\alpha$ stands 
collectively for a set of parameters. For $\alpha \ne \alpha_c$ we assume that 
$det\ W \ne 0$, so that we are dealing with a non-singular system. We assume that the singular system of interest is realized for $\alpha = \alpha_c$ (the 
subscript $c$ stands for "critical") where $det\ W(q;\alpha_c) = 0$.
Our aim is to derive the Hamilton equations of motion for the singular system, 
by taking the limit 
$\alpha \to \alpha_c$ of the Hamilton equations of motion following from 
(\ref{Lagrangean}).\footnote{Clearly the form of (\ref{Lagrangean}) is 
not unique.} For $\alpha \ne \alpha_c$, one readily constructs the Hamiltonian
\be\label{Hamiltonian}
H = \frac{1}{2}\sum_{ij}(p_i+\eta_i)W^{-1}_{ij}(p_j+\eta_j) + V(q) \ ,
\ee
where the canonical momenta are related to the velocities by
\be\label{momenta}
p_i = \sum_j W_{ij}(q;\alpha)\dot q_j - \eta_i(q) \ .
\ee
The symmetric matrix $W$ can be diagonalized by an orthogonal transformation, 
$W_D = C^TWC$
with $C_{i\ell} = v^{(\ell)}_i$, where $\vec v^{(\ell)}$ are the orthonormalized eigenvectors of $W$,
\be\label{eigenvalueeq}
W(q;\alpha)\vec v^{(\ell)}(q;\alpha) = \lambda_\ell(q;\alpha)
\vec v^{(\ell)}(q;\alpha) \ .
\ee
In terms of the eigenvalues and eigenvectors, $W$ can be written in the form
\be\label{Wmatrix}
W_{ij} = \sum_\ell \lambda_\ell v^{(\ell)}_i v^{(\ell)}_j \ .
\ee
Correspondingly $W^{-1}_{ij}$ is obtained by making the replacement 
$\lambda_\ell \to \frac{1}{\lambda_\ell}$.
Hence the Hamiltonian (\ref{Hamiltonian}) is given by
\be\label{Hamiltonian2}
H = \frac{1}{2}\sum_\ell \frac{1}{\lambda_\ell}\phi^2_\ell
+ V(q) \ ,
\ee
where 
\be\label{phi}
\phi_\ell(q,p;\alpha) := (\vec p +\vec\eta(q))\cdot\vec v^{(\ell)}(q;\alpha) \ .
\ee
For $\alpha \ne \alpha_c$ the Hamilton equations of motion then take the form
\be\label{H-eq1}
\dot q_i = \frac{\partial H}{\partial p_i} = \sum_\ell \frac{1}{\lambda_\ell}
\phi_\ell\frac{\partial\phi_\ell}{\partial p_i}\ ,
\ee
\be\label{H-eq2}
\dot p_i = -\frac{\partial H}{\partial q_i} = 
-\frac{1}{2}\sum_\ell\frac{\partial}{\partial q_i}
\left(\frac{1}{\lambda_\ell}\phi^2_\ell\right)-
\frac{\partial V}{\partial q_i}\ .
\ee
Consider now the limit $\alpha \to \alpha_c$ where $det\ W = 0$. Let $\{\lambda_{\ell_0}\}$ denote the set of eigenvalues which vanish in this limit. In order to implement the limit we first write (\ref{H-eq1}) and 
(\ref{H-eq2}) in the 
form
\be\label{q-eq1}
\dot q_i = \sum_{\ell \ne \{\ell_0\}} \frac{1}{\lambda_\ell}
\phi_\ell\frac{\partial\phi_\ell}{\partial p_i}
+ \sum_{\ell_0} \frac{1}{\lambda_{\ell_0}}
\phi_{\ell_0}\frac{\partial\phi_{\ell_0}}{\partial p_i} \ , 
\ee
\be\label{p-eq1}
\dot p_i = -\frac{1}{2}\sum_{\ell\ne \{\ell_0\}}\frac{\partial}{\partial q_i}
\left(\frac{1}{\lambda_\ell}\phi^2_\ell\right)
-\frac{1}{2}\sum_{\ell_0}\frac{\partial}{\partial q_i}
\left(\frac{1}{\lambda_{\ell_0}}\phi^2_{\ell_0}\right)-\frac{\partial V}{\partial q_i} \ .
\ee
Now, from (\ref{momenta}) and (\ref{phi}) we have that
\be\label{phiell0}
\phi_{\ell_0} = \sum_i v^{(\ell_0)}_iW_{ij}\dot q_j = 
\lambda_{\ell_0} {\vec v}^{(\ell_0)}\cdot\dot{\vec q} \ .
\ee
Hence
\be\label{rhoell01} 
\lim_{\alpha\to\alpha_c}\frac{1}{\lambda_{\ell_0}}\phi_{\ell_0} = \rho_{\ell_0}(q,\dot q;\alpha_c) \ ,
\ee
where
\be\label{rhoell02}
\rho_{\ell_0}(q,\dot q;\alpha_c) = \dot{\vec q}\cdot\vec v^{(\ell_0)}(q;\alpha_c) \ .
\ee
Note that the finiteness of the velocities in (\ref{phiell0}) implies 
that \be\label{primary}
\phi_{\ell_0}(q,p,\alpha_c) = 0 \ ,
\ee
which are just the primary constraints.
Hence for $\alpha \to \alpha_c$ (\ref{q-eq1}) reduces to
\be\label{q-eq2}
\dot q_i =  \sum_{\ell \ne \{\ell_0\}}\left(\frac{1}{\lambda_\ell}
\phi_\ell\frac{\partial\phi_\ell}{\partial p_i}\right)_{\alpha=\alpha_c} + 
\sum_{\ell_0}\rho_{\ell_0}\left(\frac{\partial\phi_{\ell_0}}
{\partial p_i}\right)_{\alpha=\alpha_c}\ .
\ee
Consider next eqs. (\ref{p-eq1}). Making again use of (\ref{phiell0}) and 
the fact that $\frac{\partial\lambda_{\ell_0}}{\partial q_i}|_{\alpha=\alpha_c} = 0 $
one obtains that for $\alpha \to \alpha_c$ these reduce to
\be\label{p-eq2}
\dot p_i =  -\frac{1}{2}\sum_{\ell \ne \{\ell_0\}}\left[\frac{\partial}{\partial q_i}
\left(\frac{1}{\lambda_\ell}\phi^2_\ell\right)\right]_{\alpha = \alpha_c} - 
\sum_{\ell_0}\rho_{\ell_0}\left(\frac{\partial\phi_{\ell_0}}{\partial q_i}\right)_{\alpha=\alpha_c}-\frac{\partial V}{\partial q_i}\ ,
\ee
where use has been made of (\ref{rhoell01}).

We now notice that the first sum on the RHS of eqs. (\ref{q-eq2}) and 
(\ref{p-eq2}) are just given by $\frac{\partial H_c}{\partial p_i}$ and 
$-\frac{\partial H_c}{\partial q_i}$, where $H_c$ is the canonical 
Hamiltonian obtained from (\ref{Hamiltonian2}) by taking the limit $\alpha\to\alpha_c$: 
\be
H_C = \frac{1}{2}\sum_{\ell\ne\ell_0}\frac{1}{\lambda_\ell(q,\alpha_c)}
\phi^2_\ell(q,p;\alpha_c) + V(q)
\ee
Here use has again be made of (\ref{primary}) and of the fact that $\lim_{\alpha\to\alpha_c}
\frac{\phi_{\ell_0}}{\lambda_{\ell_0}}$ is finite. $H_C$ is just the canonical 
Hamiltonian derived from the (\ref{Lagrangean}) for $\alpha = \alpha_c$, 
evaluated on the primary surface.
Hence the equations of motion (\ref{q-eq2}) and (\ref{p-eq2}) take the well known form 
\be\label{dotq}
\dot q_i = \frac{\partial H_T}{\partial p_i}
\ee
\be\label{dotp}
\dot p_i = -\frac{\partial H_T}{\partial q_i}
\ee
where $H_T$ is the "total" Hamiltonian 
\be\label{Htot}
H_T = H_C + \sum_{\ell_0} \rho_{\ell_0}\phi_{\ell_0} \ ,
\ee
and $\rho_{\ell_0}$ are the undetermined projections of the velocities on the zero modes (\ref{rhoell02}). Note 
that the derivatives in (\ref{dotq}) and (\ref{dotp}) are understood not to act on $\rho_{\ell_0}$. Eqs. (\ref{dotq}) and (\ref{dotp}) must be supplemented 
by the primary constraints (\ref{primary}). In fact these equations only have a solution if $q_i(t)$ and $p_i(t)$ are points 
in the submanifold defined by the primary constraints \cite{Batlle}. Note that from 
(\ref{q-eq2}) it follows that the projection of $\dot{\vec q}$ onto the 
zero mode ${\vec v}^{(\ell_0)}$ reduces to an 
identity, since $\sum_i v^{(\ell_0)}_i\frac{\partial H_c}{\partial p_i} = 0$. 
Hence we have obtained the Hamilton 
equations of motion for a constrained system in the form given by Dirac, by taking the limit 
$\alpha \to \alpha_c$ of the equations of motion for an unconstrained system. 
The primary constraints are just the statement that the projected velocities (\ref{rhoell02}) are finite in this limit.

Actually, eqs. (\ref{dotq}) 
and (\ref{dotp}), together with the primary 
constraints (\ref{primary}) do not directly yield the complete set of Lagrange equations 
of motion. These follow by also implementing the persistance in time of the 
primary constraints. The primary constraints themselves have no analog 
on the Lagrangean level but allow us to recover the connection between 
momenta and velocities needed to express the  Hamilton 
equations of motion in terms of Lagrangean variables. Thus from 
(\ref{q-eq2}) and (\ref{phi}) it follows that
\be\label{dotq3}
\dot q_i = \sum_{\ell\ne\{\ell_0\}}\frac{1}{\lambda_\ell}\phi_\ell v^{(\ell)}_i 
+ \sum_{\ell_0}(\dot{\vec q}\cdot v^{(\ell_0)})v^{(\ell_0)}_i \ ,
\ee
where it is understood that we have set  
$\alpha = \alpha_c$. Define the matrix $W_{ij}$ constructed from 
$v^{(\ell)}$ and $\lambda_\ell$ ($\ell\ne \{\ell_0\})$: 
$W_{ij} = \sum_{\ell\ne\{\ell_0\}}\lambda_\ell v^{(\ell)}_i v^{(\ell)}_j$.
From (\ref{dotq3}) we obtain  
\be\label{Wdotq}
\sum_j W_{ij}(q,\alpha_c)\dot q_j = \sum_{\ell,j} (p+\eta)_jv^{(\ell)}_iv^{(\ell)}_j
= (p+\eta)_i\ ,
\ee
where we have made use of the primary constraints (\ref{primary}), in order
to extend the sum on the RHS of (\ref{Wdotq}) over all $\ell$, and of 
the completeness relation for the eigenvectors. Hence we have recovered (\ref{momenta}) for $\alpha = \alpha_c$. 

As we have pointed out above, the persistance of the primary constraints 
yields on Lagrange level equations involving only coordinates and velocities. 
These are part of the Euler-Lagrange equations of motion. From the point 
of view taken in this paper, that the equations of motion are obtained 
by a limiting procedure from those of an unconstrained system, the 
persistance of the primary constraints can also be viewed to follow from the requirement that 
also the accelerations remain finite in the limit $\alpha\to\alpha_c$. 
Thus for $\alpha\ne\alpha_c$ we have from (\ref{phiell0}) that 
$\phi_{\ell_0} = \lambda_{\ell_0}\dot{\vec q}\cdot\vec v^{(\ell_0)}$. 
Taking the time derivative of this expression, and noting that 
$\lim_{\alpha\to\alpha_c}\lambda_{\ell_0}(q,\alpha) = 0$, and $\lim_{\alpha\to\alpha_c}\partial_i\lambda_{\ell_0}(q,\alpha) = 0$, 
we immediately conclude that $\dot\phi_{\ell_0}(q,p,\alpha_c) = 0$. On the 
Hamiltonian level this requirement must be implemented explicitely
to yield the missing Euler-Lagrange equations of motion, not manifest in (\ref{dotq}) and (\ref{dotp}). 

From the above discussion it is evident that the limit $\alpha\to\alpha_c$ 
must be carried out on the level of the Hamilton {\it equations of motion} 
of the unconstrained system, whereas on the Lagrangean level we are allowed to take this limit directly in the Lagrangean. The equivalence between the 
Lagrangean and Hamiltonian formulations has been studied in detail in \cite{Batlle}.
 
As an example consider the singular Lagrangean of the pure $U(1)$ Maxwell theory: 
\be\label{Maxwell1}
L[A^\mu,\dot A^\mu] = -\frac{1}{4}\int d^3x F_{\mu\nu}F^{\mu\nu} \ .
\ee
Consider further the non-singular Lagrangean
\be
L[A^\mu,\dot A^\mu] = -\frac{1}{4}\int d^3x F_{\mu\nu}F^{\mu\nu} + 
\frac{1}{2}\alpha\int d^3x \dot A^0 \ ,
\ee
which for $\alpha\to 0$ reduces to (\ref{Maxwell1}). This (non-covariant) choice is of course only the simplest one. Any other Lagrangean reducing to 
(\ref{Maxwell1}) in the appropriate limit would be just as acceptable. 
The canonical momenta conjugate to $A^\mu$ are given by
\be
\pi_\mu = F^{0\mu} + \alpha\delta_{\mu 0}(\partial_0 A^0) \ .
\ee
The Lagrangean written in the analogous form to (\ref{Lagrangean}) is
\be
L = \int d^3x d^3y \left[\frac{1}{2}\dot A^\mu(\vec x,x^0)
W_{\mu\nu}(\vec x,\vec y)\dot A^\nu(\vec y,x^0)\right]
-\int d^3x\ \eta_\mu(A(x))\dot A^\mu(x) - V[A] \ ,
\ee
where 
\be
\eta_\mu := \left(0,-{\vec\nabla} A^0\right) \ ,
\ee
and the matrix elements of the symmetrix matrix $W$ read 
\ba
W_{0i}(\vec x,\vec y) &=& 0 \ ,\\
W_{00}(\vec x,\vec y) &=& \alpha\delta^{(3)}(\vec x-\vec y) \ ,\\ 
W_{ij}(\vec x,\vec y) &=& \delta_{ij}\delta^{(3)}(\vec x-\vec y) \ .
\ea
The potential $V[A]$ is given by
\be
V[A] = \int d^3x \left(\frac{1}{4}F_{ij}F^{ij} - \frac{1}{2}({\vec\nabla} A^0)^2\right) \ ,
\ee
and the canonical momenta, analogous to (\ref{momenta}) read
\be
\pi_\mu(x) = \int d^3y W_{\mu\nu}(\vec x,\vec y)
\dot A^\nu(\vec y,x^0)-\eta_\mu(A(x)) \ . 
\ee
The matrix $W$ possesses the following orthonormalized eigenvectors, labeled by a discrete and continuous index, replacing the discrete index $\ell$ in (\ref{eigenvalueeq})

\be\label{eigenvectors}
v^{(\rho,\vec z)}_\nu(\vec x) = \delta_{\rho\nu}\delta^{(3)}(\vec x-\vec z) \ ,
\ee
and the corresponding eigenvalues of $W$ are given by
\be
\lambda^{(0,\vec z)} = \alpha \,;\ \  \lambda^{(i,\vec z)} = 1 \ \ (i=1,2,3)\ .
\ee
The Hamiltonian, analogous to (\ref{Hamiltonian2}) then takes the form
\ba
H &=& \frac{1}{2}\sum_i\int d^3z \left[\int d^3x \left(\pi_\mu(x) + \eta_\mu(x)\right)v^{(i,\vec z)}_\mu (\vec x)\right]^2\\
 &+&\frac{1}{2}\int d^3z \frac{1}{\alpha}\left[\int d^3x \left(\pi_\mu(x) + \eta_\mu(x)\right)v^{(0,\vec z)}_\mu (\vec x)\right]^2 + V[A]\ .
\ea
Upon making use of (\ref{eigenvectors}), this expression reduces to
\be
H = \frac{1}{2}\sum_i\int d^3z (\pi_i + \eta_i)^2 
+  \frac{1}{2}\int d^3z\frac{1}{\alpha}(\pi_0 + \eta_0)^2 + V[A] \ ,
\ee
where in the present case $\eta_0 = 0$. 
The Hamilton equations of motion read
\ba
\dot A^0 &=& \frac{\delta H}{\delta\pi_0} = \frac{\pi_0}{\alpha} \ ,\cr
\dot A^i &=&  \frac{\delta H}{\delta\pi_i} = \pi_i + \partial^iA^0 \ ,\cr
\dot\pi_0 &=& - \frac{\delta H}{\delta A^0} = -{\vec\nabla}\cdot{\vec\pi} \ ,\cr
\dot\pi_i &=&  - \frac{\delta H}{\delta A^i} = -\partial_jF^{ji} \ .
\ea
For finite $\dot A^0$ the first equation tells us that in the 
limit $\alpha\to 0$, $\pi_0$ must vanish, whereas $\dot A^0$ remains completely arbitrary. Since 
$\pi_0$ must vanish for arbitrary times, the third equation tells us that 
also $\vec\nabla\cdot\vec\pi = 0$. This is just the secondary (Gauss law) constraint. 

Alternatively we could have departed from a covariant form for an 
unconstrained system by adding to the Lagrangean density in (\ref{Maxwell1}) the covariant term 
$\frac{\alpha}{2}(\partial_\mu A^\mu)^2$. In this case $\eta_\mu(A(x))$ also depends on $\alpha$, and following our general procedure one is led to 
the equations of motion
\ba
\dot A^0 &=& \frac{\partial H}{\partial\pi_0} = 
\frac{1}{\alpha}\pi_0 - {\vec\nabla}\cdot\vec A \ ,\\
\dot A^i &=&  \frac{\partial H}{\partial\pi_i} = \pi_i +\partial^iA^0 \ ,\\
\dot \pi_0 &=&-\frac{\partial H}{\partial A^0} = -{\vec\nabla}\cdot\vec\pi \ ,\\
\dot\pi_i &=& - \partial_jF^{ji} + \partial^i\pi_0 \ .
\ea
From the first equation it follows again that in the limit $\alpha \to 0$, 
$\pi_0 = 0$ for all times, so that we are left with the standard equations for the 
pure Maxwell theory. We emphasize once more that the limit must be taken on the level of the equations of motion, and not in the Hamiltonian.
The examples demonstrate the, of course, well known fact, that only after taking into account the primary constraints and their persistance in time, the full set of Euler-Lagrange equations of motion are generated. Thus primary constraints, and possible secondary constraints following from them, play a special role, while terciary, etc. constraints correspond to consistency relations hidden in the Euler-Lagrange equations of motion. With the 
primary constraints written in the form (\ref{primary}), a 
strict iterative construction of the persistance equations for the constraints will necessarily parallel exactly the equations obtained on the Lagrangean level, irrespective of any possible bifurcations\footnote{For examples 
exhibiting bifurcations see \cite{Lusanna}}.

Let us summarize. In this paper we have shown that
the Hamilton equations of motion for any constrained system, in the form given by Dirac, 
can be obtained as a limit of the equations of motion for an unconstrained system. It was thereby shown that the primary constraints follow directly from the requirement 
that the velocities be finite in this limit. To obtain the full set of equations on Hamiltonian level, which translate into the Euler-Lagrange equations of motion, one must take into account the persistance of the primary constraints in time. These equations, which are implicit in the persistance requirement 
of the primary constraints, can also be viewed to follow from the 
requirement that also the accelerations remain finite in the above mentioned limit. The particular form of the term added to the singular Lagrangean which 
converts the system into a second class system, plays no role. The only requirement is that the unconstrained system reduces to the constrained theory of interest in an appropriate limit. We have demonstrated this for the case of 
the pure Maxwell theory.

\bigskip\noindent
{\bf Acknowledgements}

\bigskip\noindent
I thank Klaus D. Rothe for several constructive comments.

\end{document}